\documentclass[journal]{IEEEtran} 				
\IEEEoverridecommandlockouts 	
\usepackage{amsmath,amssymb}
\usepackage{amsthm}
\usepackage{color}
\usepackage[mathscr]{eucal}
\usepackage{graphics,graphicx,multicol}
\usepackage{epsfig}
\usepackage{enumerate}
\usepackage{subfigure}
\usepackage{algorithmic}
\usepackage{algorithm}
\usepackage{morefloats}
\usepackage{enumerate}
\usepackage{subfigure}
\usepackage{multirow}
\usepackage{array}
\usepackage{pstricks, pst-node, pst-plot, pst-circ}
\usepackage{moredefs}
\usepackage{cite}
\usepackage{courier}

\usepackage{epstopdf}
\DeclareGraphicsExtensions{.eps}
\usepackage{caption}
\usepackage{txfonts}
\begin{document}

\title{Application-Aware Resource Block and Power Allocation for LTE}
\author{Tugba Erpek, Ahmed Abdelhadi, and T. Charles Clancy \\
Hume Center, Virginia Tech, Arlington, VA, 22203, USA \\
\{terpek, aabdelhadi, tcc\}@vt.edu
}
\maketitle

\begin{abstract}
In this paper, we implement an application-aware scheduler that differentiates users running real-time applications and delay-tolerant applications while allocating resources. This approach ensures that the priority is given to real-time applications over delay-tolerant applications. In our system model, we include realistic channel effects of Long Term Evolution (LTE) system. Our application-aware scheduler runs in two stages, the first stage is resource block allocation and the second stage is power allocation.  In the optimal solution of resource block allocation problem, each user is inherently guaranteed a minimum Quality of Experience (QoE) while ensuring priority given to users with real-time applications. In the power allocation problem, a new power allocation method is proposed which utilizes the optimal solution of the application-aware resource block scheduling problem. As a proof of concept, we run a simulation  comparison between a conventional proportional fairness scheduler and the application-aware scheduler. The simulation results show better QoE with the application-aware scheduler. 
\end{abstract}

\pagenumbering{gobble}
\providelength{\AxesLineWidth}       \setlength{\AxesLineWidth}{0.5pt}%
\providelength{\plotwidth}           \setlength{\plotwidth}{8cm}
\providelength{\LineWidth}           \setlength{\LineWidth}{0.7pt}%
\providelength{\MarkerSize}          \setlength{\MarkerSize}{3pt}%
\newrgbcolor{GridColor}{0.8 0.8 0.8}%
\newrgbcolor{GridColor2}{0.5 0.5 0.5}%

\section{Introduction}\label{sec:intro}

The number of subscribers using wireless broadband systems and the traffic volume per subscriber are increasing every year. More smartphone users are using web-based services that require high throughput such as video streaming. With all these challenges the trend in network resource management is shifting towards using “context-awareness” in order to meet the increasing user expectations. Context-aware resource management algorithms use additional context information such as data content, type or location to make better resource allocation decisions that take the user Quality of Experience (QoE) into account. QoE is used as a measure of customer experience. 

Context-aware resource allocation is proposed in \cite{CARA} where the base station's scheduler observes context information from the user's environment and utilizes this knowledge for an efficient throughput-delay tradeoff. Similarly, a cross layer solution called Token Bucket Scheduler is proposed in \cite{Zhao} that allocates resource blocks for different applications considering the Quality of Service (QoS) requirement. Real-time applications are given priority in the proposed solution. 

An application-aware resource allocation algorithm for Long Term Evolution (LTE) downlink signal is explained and the simulation results are provided in this paper. The goal is to increase the spectrum efficiency and improve the user QoE by using the user application information in the optimization problem. As opposed to \cite{LOU} and \cite{JIANG}, the focus is not only on application types with similar throughput requirements in our proposed approach. Different applications have different performance requirements which should be fulfilled in order to provide the desired QoE for each user. With application-aware resource allocation algorithms the communication systems handle the real-time data traffic differently compared to the delay-tolerant data traffic.

In our system model and simulation, we take the realistic channel effects into account. In our simulation, we compare the application-aware algorithm performance results versus the conventional proportional fairness approach. It is shown that the application-aware approach improves the overall user QoE. Moreover, a heuristic for power control problem which also uses the application-aware scheduling results is proposed.

\subsection{Related Work}\label{sec:relatedwork}
In \cite{Perez}, the authors present a distributed approach for the joint allocation of modulation coding schemes, resource blocks, and power for LTE systems. Near optimal solutions are proposed to solve the optimization problem. In \cite{Ahmed_Utility1}-\cite{Ahmed_Utility3}, the authors present optimal rate allocation algorithms for users covered by a single carrier eNodeB. The authors use logarithmic and sigmoidal-like utility functions to represent delay-tolerant and real-time applications, respectively. In \cite{Ahmed_Utility1}, the rate allocation algorithm gives priority to real-time applications over delay-tolerant applications when allocating resources. The proposed approach is not specifically designed for LTE systems. Moreover, no channel effects are included in the results. 

In \cite{MultiBand}, a distributed solution of resource allocation for proportional fairness is provided for multi-band wireless systems. The proposed approach is not specific to the LTE systems. In \cite{SelfOrganizedLTE}, a distributed protocol that aims to achieve weighted proportional fairness by setting priority weights among UEs for LTE systems is presented. A resource block scheduling problem is formulated as a convex optimization problem. The weights play an important role while solving the optimization problem; however, the optimal resource scheduling which takes the user QoE into account is not guaranteed. All the UE applications are treated the same whenever the initial weights are set equal. A power allocation heuristic is also proposed in this paper. 

An optimal application-aware resource block scheduling is proposed in \cite{Erpek}. Initially, eNodeB makes scheduling decisions for each client assuming that the power allocations are fixed. Utility functions are used to represent the user application types and they are incorporated into the objective function while formulating the optimization problem. This approach ensures that different throughput expectations of each user are considered while making the resource block scheduling decisions.  
 
\subsection{Our Contribution}\label{sec:contribution}
Our contributions in this paper are summarized as:
\begin{itemize}
  \item We use utility functions to incorporate the user satisfaction into the resource block scheduling algorithm. This method ensures that the real-time throughput needs for each user will be met. 
  \item The optimal solution for the resource allocation problem that includes users with non-­concave utility functions (i. e. sigmoidal-like functions) and users with strictly concave utility functions (i. e. logarithmic utility functions) was proposed in \cite{Erpek} assuming that the solution to the power allocation problem is fixed. We take the findings of this paper one step further and simulate the proposed approach with realistic LTE scenario and channel effects. We compare the performance results of the application-aware approach with the weighted proportional fairness approach. 
  \item Each eNodeB needs  to allocate transmission powers on every resource block by taking the influence on the throughputs of the UEs and also the interference caused on the other UEs served by different base stations into account. Using the results of the optimal solution for application-aware resource block scheduling problem, we propose a heuristic for power allocation.
\end{itemize}

The remainder of this paper is organized as follows. Section \ref{sec:Problem_formulation} presents the system model and the problem setup. The global solution is provided in Section \ref{sec:Proof}. Our centralized resource block scheduling algorithm is presented in Section \ref{sec:Algorithm}. Section \ref{sec:power_control} provides a heuristic for the power control problem. Section \ref{sec:sim} discusses the simulation setup and provides the quantitative results of the MATLAB simulation. Section \ref{sec:conclude} concludes the paper.
\section{System Model and Problem Setup}\label{sec:Problem_formulation}

LTE standard defines a resource allocation structure in time and frequency domains. The frequency dimension is divided to subcarriers which are spaced 15 kHz apart from each other. First, the time dimension is divided to 10 ms radio frames and then into ten 1 ms subframes. Each subframe is split into two 0.5 ms slots. A resource element which is the smallest unit of resource consists of one subcarrier for a duration of one OFDM symbol. A resource block is comprised of 12 continuous subcarriers for a duration of one slot \cite{LTEBook}. 

The eNodeB decides which UE will be allocated for each resource block for a centralized resource block scheduling algorithm. A resource block can be allocated to only a single user. We use the same problem setup as in \cite{SelfOrganizedLTE} for performance comparison later in Section \ref{sec:Algorithm}. Without loss of generality, $B$ is defined to be the set of eNodeBs. $M$ denotes the set of UEs and $Z$ denotes the set of resource blocks. $z \in Z$ denotes a single resource block. $Z=F \times Q$ where each $f \in F$ represents a collection of 12 consecutive subcarriers and each $q \in Q$ represents a time slot. $r_i$ denotes the total throughput allocated by the eNodeB to the $i^{th}$ UE over all the resource blocks. Each UE has its own utility function $U_i(r_i)$ which represents the type of traffic being handled by the UE.  

\subsection{User Throughput}\label{sec:throughput}

The throughput of UE $i$ on resource block $z$ when it is scheduled by eNodeB $b(i)$ is denoted as $H_{i,b(i),z}$ and is defined as 
\begin{equation}
H_{i,b(i),z} = W \log(1+ \frac{G_{i,b(i),z}P_{b(i),z}}{N_{i,z}+\sum_{l \neq b(i)}G_{i,l,z}P_{l,z}}) 
\end{equation}
where $W$ is the bandwidth of a resource block, $N_{i,z}$ is the thermal noise experienced by UE $i$ on resource block $z$, $P_{b(i),z}$ is the transmission power that eNodeB $b(i)$ assigns to resource block $z$ and $G_{i,b(i),z}$ is the channel gain between eNodeB $b(i)$ and UE $i$ on resource block $z$. When the eNodeB $b$ transmits with power $P_{b,z}$, the received power at UE $i$ on resource block $z$ is $G_{i,b,z}P_{b,z}$. The received power, $G_{i,b,z}P_{b,z}$, of UE $i$ is considered to be its received signal strength if eNodeB $b$ is transmitting to UE $i$, and is considered to be interference, otherwise. eNodeB $b(i)$ schedules one UE in each of the resource blocks in every frame. The overall throughput of UE $i$, which is the sum of its throughput over all the resource blocks, can be written as:
\begin{equation}\label{eqn:throughput}
r_i = \sum_{z\epsilon Z}\phi_{i,b(i),z} H_{i,b(i),z} 
\end{equation}
where $\phi_{i,b(i),z}$ is the proportion of the frames that UE $i$ is scheduled by eNodeB $b(i)$ in resource block $z$.

\subsection{Utility Functions}\label{sec:utilities}

Utility functions correspond to the type of traffic being handled by the UEs. We assume all user utilities $U_i(r_i)$ in this model are strictly concave or sigmoidal-like functions. We use the same utility functions as in \cite{Ahmed_Utility1}. $U_i(0) = 0$, $U_i(r_i)$ is an increasing function of $r_i$ and $U_i(r_i)$ is twice continuously differentiable with respect to $r_i$ for the utility functions.
In our model, the normalized sigmoidal-like utility function is used, as in \cite{DL_PowerAllocation}, to represent the real-time traffic. This utility function can be expressed as 
\begin{equation}\label{eqn:sigmoid}
U_i(r_i) = c_i\Big(\frac{1}{1+e^{-a_i(r_i-b_i)}}-d_i\Big)
\end{equation}
where $c_i = \frac{1+e^{a_ib_i}}{e^{a_ib_i}}$ and $d_i = \frac{1}{1+e^{a_ib_i}}$. This utility function satisfies $U(0)=0$ and $U(\infty)=1$. Moreover, we use the normalized logarithmic utility function, as in \cite{UtilityFairness}, to represent the delay-tolerant traffic. It can be expressed as 
\begin{equation}\label{eqn:log}
U_i(r_i) = \frac{\log(1+k_ir_i)}{\log(1+k_i r_{max})}
\end{equation}
where $r_{max}$ is the required rate for the user to achieve 100\% utility percentage and $k_i$ is the rate of increase of utility percentage with the allocated rate $r_i$. This utility function also satisfies $U(0)=0$ and $U(r_{max})=1$. 
 

\subsection{Scheduling Problem}\label{sec:scheduling}
The utility proportional fairness resource scheduling problem can be formulated as:
\begin{equation}\label{eqn:opt_prob_fairness}
\begin{aligned}
& \underset{\textbf{$\phi_{i,b(i),z}$}} {\text{max}}
& & \prod_{i=1}^{M}U_i(\sum_{z\epsilon Z}\phi_{i,b(i),z} H_{i,b(i),z}) \\
& \text{subject to}
& & \sum_{i=1}^{M}\phi_{i,b(i),z} =1\\
& & &  \phi_{i,b(i),z} \geq 0, \;\;\;\;\; i = 1,2, ...,M
\end{aligned}
\end{equation}
where $M$ is the number of UEs in the coverage area of the eNodeB. The goal of this resource scheduling objective function is to allocate the resources for each UE that maximizes the total cellular network objective while ensuring proportional fairness between individual utilities. Moreover, non-zero resource allocation is guaranteed for all users. As a result, minimum QoS is ensured for all users. In addition, this approach allocates more resources to users with real-time applications which improves the QoS of the LTE system and QoE of the end user. 

\section{The Global Optimal Solution}\label{sec:Proof}

It is shown in \cite{Ahmed_Utility2} and \cite{Ahmed_Utility3} that the optimization problem (\ref{eqn:opt_prob_fairness}) is a convex optimization problem and there exists a unique tractable global optimal solution. 

As in \cite{SelfOrganizedLTE}, we use an online scheduling algorithm to decrease the computation overhead while calculating $\phi_{i,b(i),z}$. Let $\phi_{i,b(i),z}[k]$ be the proportion of the frames that the resource block $z$ is scheduled for UE $i$ in the first $k$ frames. Then, we can define the proportion of the frames that the resource block $z$ is scheduled for $i$ in the $[k+1]^{th}$ frame as:
\begin{equation*}\label{eqn:online_algorithm}
\begin{aligned}
\phi_{i,b(i),z}[k+1]=
\begin{cases}
	\frac{k-1}{k}\phi_{i,b(i),z}[k]+\frac{1}{k},\\ 
	\text{if UE $i$ is scheduled for $z$}\\
	\text{in $(k+1)^{th}$ frame}\\
	\frac{k-1}{k}\phi_{i,b(i),z}[k],\text{otherwise}.\\
\end{cases}
\end{aligned}
\end{equation*}

In this scheduling policy, the eNodeB schedules for the UE that maximizes $\frac{U'_i(\sum_{z \in Z}\phi_{i,b(i),z}H_{i,b(i),z})H_{i,b(i),z}}{U_i(\sum_{z \in Z}\phi_{i,b(i),z}H_{i,b(i),z})}$ for all $i$ and $z$ such that $\sum_{i=1}^{M}\phi_{i,b(i),z}=1$ and $\phi_{i,b(i),z} \geq 0$ \cite{Erpek}.
  

\section{Centralized Optimization Algorithm}\label{sec:Algorithm}
Our centralized resource scheduling algorithm allocates resources with utility proportional fairness. The eNodeB allocates the resource block $z$ for the UE that has the maximum $\frac{U'(\sum_{z \in Z}\phi_{i,b(i),z}H_{i,b(i),z})H_{i,b(i),z}}{ U(\sum_{z \in Z}\phi_{i,b(i),z}H_{i,b(i),z})}$. The optimization problem is solved using the utility functions. As a result, the priority will be given to the sigmoidal functions which have more strict delay and throughput requirements. 

Algorithm (\ref{alg:eNodeB}) shows our resource scheduling algorithm. This algorithm allocates resources with utility proportional fairness, which is the objective of the problem formulation. It is assumed that the application type information for each UE is received via an application programming interface. The eNodeB runs the algorithm after collecting the application type and channel gain information from each UE and makes resource scheduling decisions. The UE reference signals in the existing LTE protocol are used to estimate the channel gain.   

\begin{algorithm}
\caption{Resource Block Scheduling Algorithm}\label{alg:eNodeB}
\begin{algorithmic}
\STATE {$\phi_{i,b(i),z[k]}=0$; $r_i[k]=0$}
\FOR {$z=1 \rightarrow |Z|$}
	\STATE {Estimate the channel gain $G_{i,b(i),z}$}
	\STATE {Calculate $H_{i,b(i),z}$}
	\IF {$l = \arg \max_j\frac{U'_j(\sum_{z \in Z}\phi_{j,b(j),z}[k]H_{j,b(j),z})H_{j,b(j),z}}{U_j(\sum_{z \in Z}\phi_{j,b(j),z}[k]H_{j,b(j),z})}$} 
	\STATE {$\phi_{l,b(l),z}[k+1]=\frac{k-1}{k}\phi_{l,b(l),z}[k]+\frac{1}{k}$} \\
	\COMMENT {Resource block $z$ allocated to UE $l$}   
	\STATE {$\phi_{i,b(i),z}[k+1]=\frac{k-1}{k}\phi_{i,b(i),z}[k]$} \\
	\COMMENT {For $i \neq l$}
\ENDIF 
\ENDFOR
\end{algorithmic}
\end{algorithm}

\section{A Heuristic for the Power Control Problem}\label{sec:power_control}
In this section, the power control problem is discussed, i.e., how the eNodeBs choose $P_{b,z}$ in order to solve the optimization problem (\ref{eqn:opt_prob_fairness}). The problem setup and assumptions are again defined as given in \cite{SelfOrganizedLTE}. The eNodeBs need to know the solution of the scheduling problem $\phi_{i,b(i),z}$ and the values of channel gains $G_{i,b(i),z}$ in order to choose suitable power values. To reduce computation and communication overhead, the eNodeBs assume that for all clients $i$ associated with eNodeB $b$, the perceived thermal noises are all $N_{b,z}$, the channel gains between them and eNodeB $b$ are all $G_{b,b,z}$, and the channel gains between them and eNodeB $l$ ($l\neq b$) are all $G_{b,l,z}$  on resource block $z$. With these assumptions the power control problem can be written as:
\begin{equation}\label{eqn:PowerControl}
\begin{aligned}
& {\text{max}}\sum_{b}\sum_{i \in I}\log(U_i(\sum_{z\epsilon Z}\phi_{i,b(i),z}^* B \log(1+\text{SINR}_{b,z}(P)))) \\
& \text{s.t.}\sum_{f \in F}{P_{b,(f,q)} \leq W}, \forall b \in B, q \in Q, P_{b,z} \geq 0, \forall i,b,z \\
\end{aligned}
\end{equation}
where $\phi_{i,b(i),z}^*$ is the solution to the scheduling problem, thus known values and 
\begin{equation*}\label{eqn:SINRbz}
\begin{aligned}
& {\text{SINR}}_{b,z}(P)=\frac{G_{b,b,z}P_{b,z}}{N_{b,z}+\sum_{l \neq b} G_{b,l,z}P_{l,z}}
\end{aligned}
\end{equation*}
where $P$ is the vector consisting of $\{P_{b,z}\}$.
This problem is non-convex. A distributed heuristic that converges to a local optimum is proposed instead. A gradient method is applied to solve this problem. 
\begin{equation}\label{eqn:Y_P1}
\begin{aligned}
Y(P) := \sum_b \sum_{i \in I} \log(U_i(\sum_{z \in Z}\phi_{i,b(i),z}^*B\log(1+\text{SINR}_{b,z}(P))))
\end{aligned}
\end{equation}
Then we have:	
\begin{equation}\label{eqn:Y_P2}
\begin{aligned}
& Y(P) := \sum_{i \in I}\log(U_i(\sum_{z \in Z}\phi_{i,b(i),z}^*B\log(1+\text{SINR}_{b,z}(P))))
\\
& +\sum_{o \neq b}\sum_{i \in I}\log(U_i(\sum_{z \in Z}\phi_{i,o(i),z}^*B\log(1+\text{SINR}_{o,z}(P))))
\end{aligned}
\end{equation}
where 
\begin{equation*}\label{eqn:SINR2}
\begin{aligned}
\text{SINR}_{o,z}(P) := \frac{G_{o,o,z}P_{o,z}}{N_{o,z}+\sum_{\substack{l \neq o \\ l \neq b}} G_{o,l,z}P_{o,z}+G_{o,b,z}P_{b,z}}
\end{aligned}
\end{equation*}
and
\begin{equation}\label{eqn:deltaY}
\begin{aligned}
& \frac{\partial Y(P)}{\partial P_{b,z}}=\sum_i\frac{U_i^\prime(\sum_{z \in Z}\phi_{i,b(i),z}^* B \log(1+\text{SINR}_{b,z}(P)))}{U_i(\sum_{z \in Z}
\phi_{i,b(i),z}^* B \log(1+\text{SINR}_{b,z}(P)))} \\
& \phi_{i,b(i),z}^* B\frac{G_{b,b,z}}{N_{b,z}+\sum_lG_{b,l,z}P_{l,z}} + \\
& \sum_{o \neq b}\sum_i \frac{U_i^\prime(\sum_{y \in Z}\phi_{i,o(i),y}^*B\log(1+\text{SINR}_{o,y}(P)))}{U_i(\sum_{y \in Z}\phi_{i,o(i),y}^* B
\log(1+\text{SINR}_{o,y}(P)))} \\
& \phi_{i,o(i),z}^* B\frac{-G_{o,o,z}P_{o,z}G_{o,b,z}}{(N_{o,z}+\sum_l G_{o,l,z}P_{l,z})(N_{o,z}+\sum_{l \neq o}G_{o,l,z}P_{l,z})}
\end{aligned}
\end{equation}

Each eNodeB updates its power periodically. When eNodeB $b$ updates its power, it sets its power on resource block $(f,q)$ to be:
\begin{equation}\label{eqn:deltaY}
\begin{aligned}
\begin{cases}
    \big\{P_{b,(f,q)}+\alpha\frac{\partial Y(P)}{\partial P_{b,
(f,q)}}\big\}^+, \text{if} \sum_e[P_{b,(e,q)}+\alpha\frac{\partial Y(P)}{\partial P_{b,(e,q)}}]^+ \leq W, \\
\\
    W\frac{[P_{b,(f,q)}+\alpha\frac{\partial Y(P)}{\partial P_{b,
(f,q)}}\big]^+}{\sum_e[P_{b,(e,q)}+\alpha\frac{\partial Y(P)}{\partial P_{b,(e,q)}}]^+}, \text{otherwise},
  \end{cases}
\end{aligned}
\end{equation}
where $x^+ := \text{max}\{x,0\}$ and $\alpha$ is a small constant. The calculated power is more throughput efficient since $\phi_{i,b(i),z}^*$ terms are calculated using application-aware approach  compared to the solution proposed in \cite{SelfOrganizedLTE}.

\section{Simulation Results}\label{sec:sim}

In this section, we present and compare the simulation results for both the application-aware resource scheduling and conventional proportional fairness algorithms \cite{SelfOrganizedLTE}. 

An LTE system with one eNodeB and six UEs are considered in the simulation. It is assumed that there are 200 resource blocks to be scheduled. The simulation area is 500m x 500m. The eNodeB is placed in the middle of the simulation area and six UEs are uniformly distributed.

Channel gains are derived using the following equation:
\begin{equation}\label{eqn:ChGain}
\begin{aligned}
\text{PL}(d)=128.1+37.6\log_{10}(d)+X+Y 
\end{aligned}
\end{equation}
where $\text{PL}(d)$ is the channel gain in dB and $d$ is distance in km. $X$ and $Y$ represent shadowing and fast fading, respectively. $X$ is the log-normal shadowing with mean 0 dB and standard deviation 8 dB. $X$ represents slow fading; as a result, it is considered that it is time invariant. However, it varies in frequency, in every 180 kHz. $Y$ represents Rayleigh fast fading with 5 Hz Doppler frequency. Therefore, it also varies in frequency. It is assumed that the thermal noise is $3.5\times10^{-15} W$ for all the UEs \cite{SelfOrganizedLTE}.

\begin{figure}
    \centering
    \includegraphics[width=3.5in]{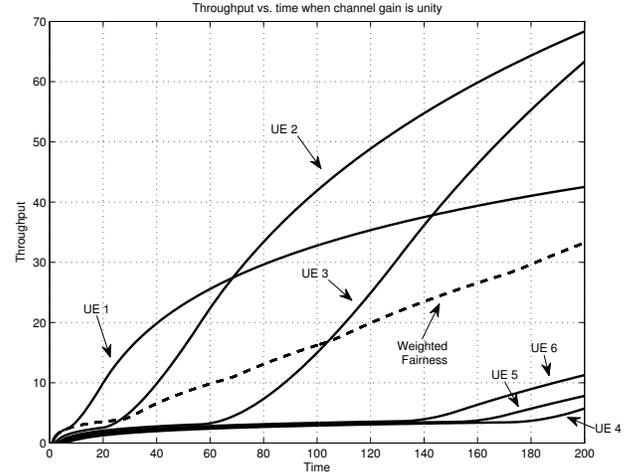}
    \captionsetup{justification=centering}
    \caption{Throughput for each user when the channel gain is unity.}
    \label{fig:sim:UnityGain}
\end{figure}

It is assumed that the first three users have sigmoidal-like utility functions. $a=5$, $b=10$ for the first user which is an approximation to a step function (e.g. VoIP), $a=3$, $b=20$ for the second user which is an approximation of an adaptive real-time application (e.g. standard definition video streaming), and $a=1$, $b=30$ for the third user which is also an approximation of an adaptive real-time application (e.g. high definition video streaming). Three logarithmic functions are used for the last three users with $r_{max}=100$ and different $k_i$ parameters which are approximations for delay tolerant applications (e.g. FTP). $k={15, 3, 0.5}$ are used, respectively. The simulation was run in MATLAB. Algorithm (\ref{alg:eNodeB}) is used while making resource block scheduling decisions. Initially, unity channel gain is assumed in order to verify the simulation. The expectation is UE 1 will initially get the most of the resources followed by UE 2 and UE 3. The throughput levels for the UEs with logarithmic utility functions will increase more when the first three users are satisfied with the rate they are allocated. Figure \ref{fig:sim:UnityGain} shows the results. As expected, the throughput of the applications with sigmoidal-like utility functions are higher compared to the logarithmic utility functions. Note that a minimum
resource allocation for all users independent of their application type is guaranteed. 

The weighted proportional fairness algorithm proposed in \cite{SelfOrganizedLTE} is also simulated for comparison purposes. Priority weights, $w_i$, which are user-dependent priority indicators are used in \cite{SelfOrganizedLTE} for each UE. The weighted proportional fairness is achieved by scheduling the resource block to the UE which maximizes $\frac{w_i H_{i,b(i),z}}{r_i}$. The priority weights are set as unity in the simulation. As a result, it is expected that all users are treated equally which is the same as the conventional proportional fairness approach. The results of proportional fairness algorithm are shown with a dashed line in Figure \ref{fig:sim:UnityGain}. The UEs are not differentiated based on the traffic type and they all have the same throughput. Even though the rate for users with elastic traffic is initially higher with proportional fairness approach, this does not increase the user QoE since elastic traffic can easily adapt to the network conditions.  
 
\begin{figure}
    \centering
    \includegraphics[width=3.5in]{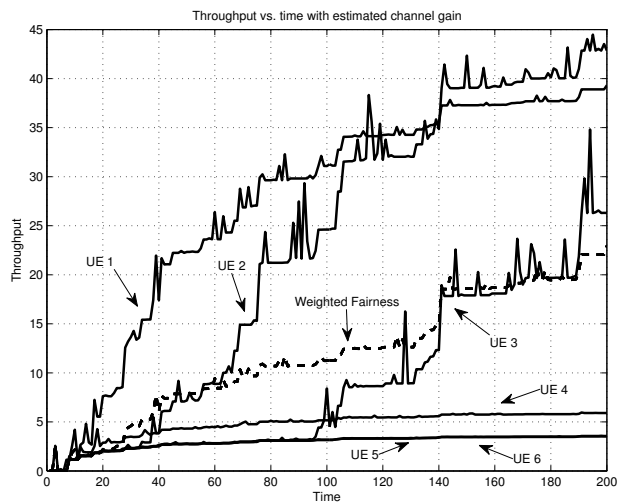}
    \captionsetup{justification=centering}
    \caption{Throughput for each user with estimated channel gain.}
    \label{fig:sim:GainIncluded}
\end{figure}

Next, channel gain is introduced to the simulation. Path loss values are calculated for each user before assigning a resource block and throughput per resource block per user is calculated. The simulation results are averaged over 200 iterations. Figure \ref{fig:sim:GainIncluded} shows the results. The curves are not smooth anymore. However, it can be observed that the overall pattern is the same and the application-aware resource allocation approach takes the traffic type into account while making decisions.

\section{Conclusion}\label{sec:conclude}
An application-aware resource block scheduling algorithm was presented in this paper. A simulation was developed with users having different QoE expectations. The realistic channel effects were taken into account for the simulation and the throughput results for each user were compared with the proportional fairness approach. It was shown that the application-aware algorithm increases the overall QoE by giving priority to users with real-time applications over delay-tolerant ones. Furthermore, a heuristic for the power control problem was proposed in the paper. The power control algorithm uses the resource allocation decisions from the application-aware approach.
\bibliographystyle{ieeetr}
\bibliography{pubs}

\begin{thebibliography}{10}

\bibitem{CARA}
M.~Proebster, M.~Kaschub, T.~Werthmann, and S.~Valentin, ``Context-aware
  resource allocation for cellular wireless networks,'' in {\em IEEE Wireless
  Communications Magazine}, February 2014.

\bibitem{Zhao}
L.~Zhao, Y.~Qin, M.~Ma, X.~Zhong, and L.~Li, ``{QoS} guaranteed resource block
  allocation algorithm in {LTE} downlink,'' in {\em International ICST
  Conference on Communications and Networking in China}, 2012.

\bibitem{LOU}
C.~Lou and L.~Qiu, ``{QoS}-aware scheduling and resource allocations for video
  streams in e-{MBMS} towards lte-a system.,'' {\em Proc. of the IEEE Vehicular
  Technology Conference}, September 2011.

\bibitem{JIANG}
D.~Jiang, H.~Wang, E.~Malkamaki, and E.~Tuomaala, ``Principle and performance
  of semi-persistent scheduling for {VoIP} in {LTE} system,'' {\em Proc. of the
  International Conference on Wireless Communications, Networking and Mobile
  Computing}, September 2007.

\bibitem{Perez}
D.~L\'{o}pez-P\'{e}rez, \'{A}kos Lad\'{a}nyi, H.~R. Alp\'{a}r~J\"{u}ttner, and
  J.~Zhang, ``Optimization method for the joint allocation of modulation
  schemes, coding rates, resource blocks and power in self-organizing {LTE}
  networks,'' April 2011.

\bibitem{Ahmed_Utility1}
A.~Abdel-Hadi and C.~Clancy, ``A utility proportional fairness approach for
  resource allocation in {4G-LTE},'' in {\em IEEE International Conference on
  Computing, Networking and Communications: Computing, Networking and
  Communications Symposium (ICNC'14 - CNC)}, 2014.

\bibitem{Ahmed_Utility3}
A.~Abdel-Hadi, C.~Clancy, and J.~Mitola, ``A resource allocation algorithm for
  multi-application users in {4G-LTE},'' in {\em ACM Proceedings of the
  19$^{\text{th}}$ Annual International Conference on Mobile Computing and
  Networking (MobiCom)}, 2013.

\bibitem{MultiBand}
I.-H. Hou and P.~Gupta, ``Distributed resource allocation for proportional
  fairness in multi-band wireless systems,'' in {\em IEEE International
  Symposium on Information Theory Proceedings}, 2011.

\bibitem{SelfOrganizedLTE}
I.-H. Hou and C.~S. Chen, ``Self-organized resource allocation in {LTE} systems
  with weighted proportional fairness,'' in {\em IEEE International Conference
  in Communications (ICC)}, 2012.

\bibitem{Erpek}
T.~Erpek, A.~Abdelhadi, and C.~Clancy, ``An optimal application-aware resource
  block scheduling in {LTE},'' February 2015.

\bibitem{LTEBook}
S.~Sesia, I.~Toufik, and M.~Baker, {\em LTE - The UMTS Long Term Evolution:
  From Theory to Practice}.
\newblock John Wiley Son, 2011.

\bibitem{DL_PowerAllocation}
J.-W. Lee, R.~R. Mazumdar, and N.~B. Shroff, ``Downlink power allocation for
  multi-class wireless systems,'' {\em IEEE/ACM Trans. Netw.}, vol.~13,
  pp.~854--867, Aug. 2005.

\bibitem{UtilityFairness}
G.~Tychogiorgos, A.~Gkelias, and K.~K. Leung, ``Utility-proportional fairness
  in wireless networks.,'' in {\em PIMRC}, pp.~839--844, IEEE, 2012.

\bibitem{Ahmed_Utility2}
A.~Abdel-Hadi and C.~Clancy, ``A robust optimal rate allocation algorithm and
  pricing policy for hybrid traffic in {4G-LTE},'' in {\em IEEE
  24$^{\text{th}}$ International Symposium on Personal, Indoor and Mobile Radio
  Communications (PIMRC)}, 2013.

\end{thebibliography}
\end{document}